\documentclass[aps, prb, twocolumn, amsmath, amssymb, showpacs]{revtex4-1}

\usepackage[utf8]{inputenc}
\usepackage{graphicx}
\usepackage{units}
\usepackage{color}
\usepackage[pdftex, colorlinks=true, linkcolor=myblue, citecolor=myblue,
urlcolor=myblue]{hyperref}

\definecolor{myblue}{rgb}{0,0,1}

\begin{document}

\title{Cotunneling, current blockade, and backaction forces in
nanobeams close \\ to the Euler buckling instability}

\author{Guillaume Weick}
\email{Guillaume.Weick@ipcms.unistra.fr}
\author{Dominique M.-A. Meyer}
\affiliation{Institut de Physique et Chimie des Mat\'eriaux de Strasbourg (UMR
7504), CNRS and Universit\'e de Strasbourg, 23 rue du Loess, BP 43, F-67034
Strasbourg Cedex 2, France}


\begin{abstract}
Single-electron transistors embedded in a vibrating nanoresonator such as a
doubly-clamped carbon nanotube exhibit effects stemming from the
coupling between electronic and vibrational degrees of freedom. In particular, a
capacitive electromechanical coupling induces a blockade of the current at low
bias voltage. It has been recently shown theoretically
within a 
sequential-tunneling approximation
that this current blockade can be enhanced by orders of 
magnitude when the suspended structure is
brought to the Euler buckling instability. Here, we investigate the role of
cotunneling on the predicted enhancement and 
show that the latter is not suppressed by cotunneling effects. We further demonstrate that despite the fact that the 
current blockade is difficult to measure far from the Euler instability, the
backaction of the current flow on the nanobeam
frequency may be easier to observe.
\end{abstract}


\maketitle

\section{Introduction}
Nanoelectromechanical systems \cite{craig00_Science, rouke01_PhysWorld,
ekinc05_RSI, poot11_preprint} 
exhibit effects originating from the coupling between electronic and vibrational
degrees of freedom that go beyond the ones
encountered in more conventional nanostructures such as quantum dots. 
A prominent example of these new effects is the low-bias current blockade induced by a
capacitive electromechanical coupling when a single-electron transistor is
embedded in a vibrating structure. \cite{pisto07_PRB} This phenomenon is the
classical counterpart of the Franck-Condon blockade in molecular devices
\cite{koch05_PRL, koch06_PRB} that has been observed in suspended carbon
nanotubes for high-energy vibrational modes. \cite{letur09_NaturePhysics}
For low-energy (classical) vibrations, the blockade is difficult to observe due to the
relatively weak electromechanical coupling typically encountered in experiments.
However, it has been reported that this capacitive coupling is
responsible for the modification of the flexural vibration frequency 
when electrons tunnel on a suspended carbon nanotube.
\cite{steel09_Science, lassa09_Science}

\begin{figure}[tb]
\includegraphics[width=.95\columnwidth]{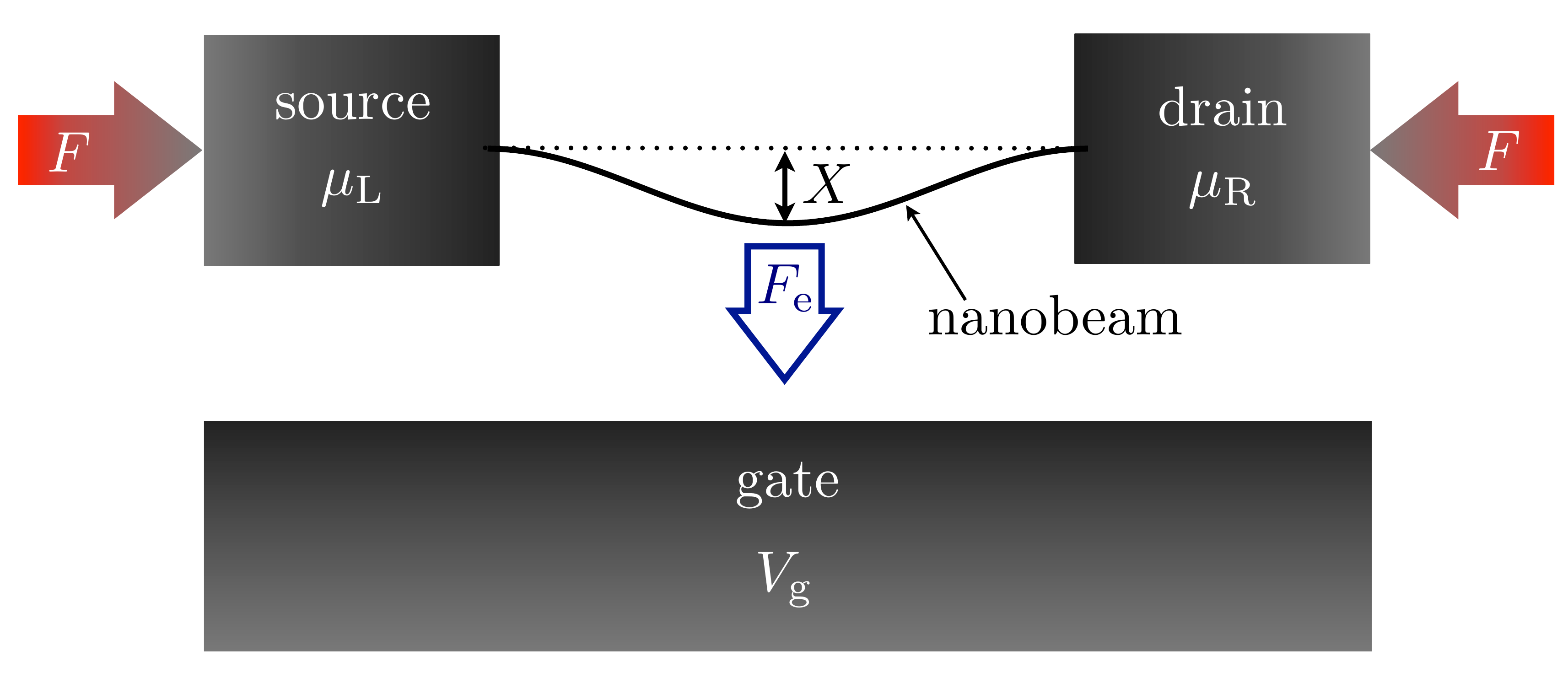}
\caption{
\label{fig:setup}%
(Color online)
Sketch of the setup: a suspended doubly-clamped nanobeam
(e.g., a carbon nanotube) with
deflection amplitude $X$ forming a quantum dot connected 
via tunnel barriers to source and drain electron reservoirs held at chemical 
potentials $\mu_\mathrm{L}$ and $\mu_\mathrm{R}$ by the bias voltage 
$V$, respectively. The nanobeam is capacitively coupled to a metallic electrode 
kept at a gate voltage $V_\mathrm{g}$, which induces a stochastic force
$F_\mathrm{e}$ that attracts the beam
towards the gate electrode whenever the quantum dot is charged.
The additional compression force $F$ acts on the nanobeam
and induces the buckling instability.}
\end{figure}

In Ref.~\onlinecite{weick11_PRB}, a way of enhancing the classical current blockade by
orders of magnitude by exploiting the well-known Euler buckling instability
\cite{landau} has been
recently proposed:
The energy scale associated with the bias voltage below which transport is
blocked is given by \cite{pisto07_PRB}
$E_\mathrm{E}=F_\mathrm{e}^2/m\omega^2$, where $F_\mathrm{e}$ is the electromechanical
coupling resulting from the gate capacitance of the device and $\omega$ and $m$ the
vibrational frequency and mass of its mechanical part, respectively.
When a lateral compressive force $F$ is exerted on a nanobeam forming a quantum
dot coupled through tunnel barriers to source and drain reservoirs and capacitively 
coupled to a gate electrode as sketched in
Fig.~\ref{fig:setup}, its vibrational frequency
$\omega$ goes down to zero when $F$ reaches the critical force $F_\mathrm{c}$ at
which buckling occurs. \cite{landau} Thus, one expects a large enhancement of the bias window
below which current through the device is suppressed. The argument above would
in principle suggest that the energy scale $E_\mathrm{E}$ diverges at the instability.
However, anharmonicities in the vibrations of the nanobeam, which become crucial close to the
Euler instability, cut off the apparent divergence. \cite{weick11_PRB}

First studied by Leonhard Euler in 1744 while investigating the maximal load
that a column can sustain, \cite{euler} the buckling instability is a paradigm for a continuous mechanical
instability. \cite{landau} It has been the subject of various experimental
studies in micro- and nanomechanical systems over the past years.
\cite{falvo97_Nature, ponch99_Science, carr03_APL, carr05_EPL, roode09_APL} In
particular, buckling instabilities have been observed in carbon nanotubes \cite{falvo97_Nature,
ponch99_Science}
and in SiO$_2$ nanobeams \cite{carr03_APL, carr05_EPL} where it was demonstrated
that, despite the small size of these
objects, the mechanical instability can be explained in terms of continuous
classical elasticity theory. Micromechanical beams beyond the Euler
instability were also proposed to serve as mechanical memories in
Ref.~\onlinecite{roode09_APL}.
The quantum properties of nanobeams close to the
Euler instability were theoretically studied in
Refs.~\onlinecite{carr01_PRB}--\onlinecite{savel07_PRB}. 
In particular, these systems were proposed
to investigate zero-point fluctuations of a mechanical mode \cite{werne04_EPL}
or to serve as a mechanical qubit. \cite{savel06_NJP, savel07_PRB}
The Euler instability has been recently theoretically considered in
nano\textit{electro}mechanical systems
\cite{weick10_PRB} in the
case where the electromechanical coupling $F_\mathrm{e}$ is negligible with
respect to the intrinsic electron-phonon coupling. \cite{maria09_PRB} It was
shown in Ref.\ \onlinecite{weick10_PRB} that
this intrinsic coupling changes qualitatively the nature of the instability,
turning it into a discontinuous one. \cite{savel04_PRB}

The results of Ref.~\onlinecite{weick11_PRB} where the enhancement of the current blockade at the Euler
instability was predicted were obtained in the 
sequential-tunneling regime of weak coupling to the electron reservoirs,
i.e., $\hbar\Gamma\ll k_\mathrm{B}T$ with $\Gamma$ the typical tunneling-induced
width (or hybridization) of the dot energy levels and $T$ the temperature, 
where cotunneling currents are exponentially suppressed. In this paper, 
we investigate the role played by cotunneling that
dominates the current inside the Coulomb diamond and which becomes relevant
in the opposite regime of strong coupling to the leads, i.e., $\hbar\Gamma\gg
k_\mathrm{B}T$. 
We show that the current blockade survives for
hybridization to the leads smaller than the energy scale $E_\mathrm{E}$ set by the bias voltage
below which current is suppressed. 
Moreover, we show that the backaction of the current flow on the mechanical part
of the device leads to large current-induced frequency shifts of the
nanobeam that may be easier to observe experimentally than the current blockade
when the nanobeam is far from the mechanical instability.

The paper is organized as follows: We start by presenting our model in Sec.~\ref{sec:model}. 
Section \ref{sec:FP} describes the effective statistical
description in terms of a current-induced Langevin process of the nanobeam that we use to solve the model. 
Results for the transport characteristics of the device are given in Sec.~\ref{sec:IV},
while Sec.~\ref{sec:backaction} presents our predictions concerning the
backaction that the current flow exerts on the nanoresonator. We finally
conclude in Sec.~\ref{sec:ccl}.

\section{Model}
\label{sec:model}
The model we consider is sketched in Fig.~\ref{fig:setup}:
A suspended nanobeam forms a quantum dot connected to source and drain leads via tunnel
barriers. A
lateral compression force $F$ is exerted on the nanotube and brings it
to the Euler buckling instability when $F$ reaches
the critical force $F_\mathrm{c}$. \cite{landau} The 
total Hamiltonian of the system reads
\begin{equation}
\label{eq:H}
H=H_\mathrm{vib}+H_\mathrm{SET}+H_\mathrm{c}, 
\end{equation}
where $H_\mathrm{vib}$ describes the flexural vibrations of the nanobeam,
$H_\mathrm{SET}$ accounts for the electronic degrees of freedom of the
single-electron transistor (SET) and $H_\mathrm{c}$ represents the coupling
between the mechanical motion of the beam and the electrons residing on the
quantum dot.

At sufficiently low temperature, only the lowest-energy bending mode of amplitude
$X$ is
significantly populated, and the resulting vibrational Hamiltonian in Eq.~\eqref{eq:H}
takes the Landau-Ginzburg form
\cite{weick11_PRB, carr01_PRB, werne04_EPL, peano06_NJP, savel06_NJP,
savel07_PRB, weick10_PRB}
\begin{equation}
\label{eq:H_vib}
H_\mathrm{vib}=\frac{P^2}{2m}+\frac{m\omega^2}{2}X^2+\frac{\alpha}{4} X^4,
\end{equation}
with $P$ the conjugated momentum to $X$ and $m$ the mass of the mode.
In Eq.~\eqref{eq:H_vib}, 
\begin{equation}
\label{eq:omega}
\omega=\omega_0\sqrt{1-\frac{F}{F_\mathrm{c}}}
\end{equation}
denotes the compression force-dependent vibrational frequency of the fundamental bending
mode of the beam, with $\omega_0$ the corresponding frequency for vanishing
compression force $F$.
The anharmonicity parameter $\alpha>0$ stabilizes the mode for
$F>F_\mathrm{c}$ ($\omega^2<0$) where the
nanobeam buckles in one of the two metastable positions at
$X=\pm\sqrt{-m\omega^2/\alpha}$. For $F<F_\mathrm{c}$ ($\omega^2>0$), the
nanobeam stays flat at $X=0$. For clamped boundary conditions, one can express
\cite{weick11_PRB} $F_\mathrm{c}=\kappa(2\pi/L)^2$,
$\omega_0=\sqrt{\kappa/\sigma}(2\pi/L)^2$, $m=3\sigma L/8$ and
$\alpha=F_\mathrm{c}L(\pi/2L)^4$ in terms of the bending rigidity $\kappa$, the
length $L$ and the linear mass density $\sigma$ of the beam.

The single-electron transistor is modelled by
the Hamiltonian 
\begin{equation}
H_\mathrm{SET}=H_\mathrm{dot}+H_\mathrm{leads}+H_\mathrm{tun},
\end{equation}
with 
\begin{equation}
\label{eq:H_dot}
H_\mathrm{dot}=(\epsilon_\mathrm{d}-e\bar
V_\mathrm{g})n_\mathrm{d}
\end{equation}
the Hamiltonian describing the single-level quantum dot of energy 
$\epsilon_\mathrm{d}$, where
$\bar V_\mathrm{g}=C_\mathrm{g}V_\mathrm{g}/C_\Sigma$ is
the (effective) gate voltage which is expressed in terms of the applied gate
voltage $V_\mathrm{g}$, the capacitance of the gate electrode $C_\mathrm{g}$ and
the total capacitance $C_\Sigma$ of the device.
The occupation operator
reads $n_\mathrm{d}=d^\dagger d$ where $d^\dagger$ ($d$) creates (annihilates) 
an electron on the dot. 
We consider for simplicity spinless electrons, as the inclusion of spin as well as
onsite Coulomb repulsion in Eq.~\eqref{eq:H_dot} should not qualitatively change our results. 
The electrons in the left (L) and right (R) reservoirs are considered to
constitute Fermi liquids. The corresponding Hamiltonian reads
\begin{equation}
H_\mathrm{leads}=\sum_{ka}(\epsilon_k-\mu_a)c^\dagger_{ka}c^{\phantom{\dagger}}_{ka}, 
\end{equation}
where $c_{ka}^\dagger$ ($c_{ka}$) creates 
(annihilates) an electron of momentum $k$ with energy $\epsilon_k$ in lead
$a=\mathrm{L,R}$.
The lead $a$ is maintained at the chemical potential $\mu_a$ by the bias voltage $V$, 
such that $\mu_\mathrm{L}-\mu_\mathrm{R}=eV$. In what follows, we assume the
voltage drop across the junction to be symmetric, i.e., 
$\mu_\mathrm{L}=-\mu_\mathrm{R}=eV/2$.
Tunneling between the quantum dot and the left and right leads is accounted
for by the Hamiltonian
\begin{equation}
H_\mathrm{tun}=\sum_{ka}(t_a
c^\dagger_{ka}d+t_a^*d^\dagger c_{ka}),
\end{equation}
with $t_a$ the tunneling amplitude between the dot and lead $a$. 

The flexural vibrations of the nanobeam
couple to electronic transport due to the gate electrode that exerts the
electrostatic force $-F_\mathrm{e}$ when the quantum dot is charged with an extra
electron \cite{steel09_Science, lassa09_Science, sapma03_PRB} (see
Fig.~\ref{fig:setup}). 
The resulting coupling Hamiltonian, linear in the deflection $X$, reads
\begin{equation}
\label{eq:H_c}
H_\mathrm{c}=F_\mathrm{e}X n_\mathrm{d}.
\end{equation}
Since the occupation of the quantum dot $n_\mathrm{d}$ (which varies in our
model between 0 and 1) follows a Poissonian
statistics, the coupling \eqref{eq:H_c} results in a stochastic force which is
exerted on the nanotube whenever the latter is charged with an extra electron. 
Such a coupling induces a blockade of the current at low bias voltage
\cite{pisto07_PRB} which can be dramatically enhanced at the Euler buckling
instability. \cite{weick11_PRB} Due to the backaction that the current flow exerts on
the nanobeam, the coupling \eqref{eq:H_c} further induces a renormalization of
its vibrational frequency, as recently observed in suspended carbon nanotubes.
\cite{steel09_Science, lassa09_Science} For quantized high-energy vibrational modes, a linear
electron-vibron coupling similar to Eq.\ \eqref{eq:H_c}, but not induced by the gate
capacitance, leads to the Franck-Condon blockade.
\cite{koch05_PRL, koch06_PRB, letur09_NaturePhysics}

On top of the electrostatic potential \eqref{eq:H_c}, 
a second kind of coupling between the mechanical and electronic degrees of
freedom stems from the electron-phonon coupling intrinsic
to the tube and is quadratic in $X$. \cite{maria09_PRB} As shown in
Ref.~\onlinecite{weick10_PRB}, such a coupling leads to a current-induced discontinuous
Euler instability and to a novel mechanism of current blockade at
low bias voltage, termed ``tricritical current blockade". 
The relative importance of the two types of coupling
mentioned above is
controlled by the distance $h$ between the nanobeam and the gate electrode. Indeed,
the intrinsic coupling does not depend on $h$, while the strength of the
gate-induced coupling \eqref{eq:H_c} depends logarithmically on $h$.
\cite{sapma03_PRB} In the
remainder of this paper, we assume that the distance $h$ is sufficiently small
such that the coupling \eqref{eq:H_c} dominates and thus neglect the intrinsic
electron-phonon coupling.

\section{Current-induced Langevin dynamics}
\label{sec:FP}
The model encapsulated in the Hamiltonian \eqref{eq:H} can be solved in an essentially
exact way by noticing that close to the Euler instability, where
$\omega\rightarrow0$ [see Eq.~\eqref{eq:omega}], (i) the vibrational mode is classical, $\hbar\omega\ll
k_\mathrm{B}T$ with $T$ the temperature, and (ii) it is extremely slow as
compared to the electronic dynamics, $\omega\ll\Gamma$, where $\Gamma$ is the
typical tunneling-induced width (or hybridization) of the energy level of the
dot. Within a
non-equilibrium Born-Oppenheimer approximation
\cite{mozyr06_PRB, pisto08_PRB, husse10_PRB, nocer11_PRB, bode11_PRL} which exploits the separation of
timescales between fast electronic and slow vibrational degrees of freedom, and 
which is asymptotically exact at the buckling instability where
$\omega\rightarrow0$,
\cite{weick11_PRB, weick10_PRB} the dynamics of the nanobeam is described by a Langevin
process with Gaussian white noise, equivalent \cite{zwanzig} to the Fokker-Planck equation 
\begin{align}
\label{eq:FP}
\partial_t\mathcal{P}=&-\frac{P}{m}\partial_X\mathcal{P}-F_\mathrm{eff}(X)\partial_P\mathcal{P}
\nonumber\\
&+\frac{\eta(X)}{m}\partial_P(P\mathcal{P})+\frac{D(X)}{2}\partial_P^2\mathcal{P}
\end{align}
for the probability $\mathcal{P}(X, P, t)$ to find the resonator at phase-space
point $(X, P)$ at time $t$. The effective force
\begin{equation}
\label{eq:F_eff}
F_\mathrm{eff}(X)=-m\omega^2X-\alpha X^3-F_\mathrm{e}n_0(X)
\end{equation}
can be easily inferred from
averaging the Heisenberg equations of motion for $X$ and $P$ on a timescale
long as compared to $\Gamma^{-1}$, but short relative to $\omega^{-1}$. In
Eq.\ \eqref{eq:F_eff}, 
the term proportional to the average dot occupation for fixed $X$,
$n_0(X)=\langle n_\mathrm{d}\rangle_X$, corresponds to a conservative force that
modifies the dynamical properties of the nanobeam in presence of a current flow.
In
Eq.\ \eqref{eq:FP}, $\eta(X)$ and $D(X)$ represent a current-induced
friction and fluctuation coefficient, respectively. 
These quantities can be
expressed through the non-equilibrium lesser and
greater Green's functions of the dot at fixed $X$, 
$G_\mathrm{d}^<(t, X)=\mathrm{i}\langle d^\dagger d(t) \rangle_X$ and 
$G_\mathrm{d}^>(t, X)=-\mathrm{i}\langle d(t) d^\dagger \rangle_X$,
respectively.
One finds \cite{mozyr06_PRB, pisto08_PRB, husse10_PRB, nocer11_PRB, bode11_PRL}
\begin{subequations}
\begin{align}
n_0(X)&=-\mathrm{i}\int\frac{\mathrm{d}\epsilon}{2\pi\hbar}G_\mathrm{d}^<(\epsilon,
X),\\
\eta(X)&=F_\mathrm{e}^2\int\frac{\mathrm{d}\epsilon}{2\pi\hbar}G_\mathrm{d}^<(\epsilon, X)
\partial_\epsilon G_\mathrm{d}^>(\epsilon, X),\\
D(X)&=F_\mathrm{e}^2\int\frac{\mathrm{d}\epsilon}{2\pi\hbar}G_\mathrm{d}^<(\epsilon, X)
G_\mathrm{d}^>(\epsilon, X),
\end{align}
\end{subequations}
with 
\begin{subequations}
\label{eq:GF}
\begin{align}
G_\mathrm{d}^<(\epsilon, X)&=
\int\mathrm{d}t\,\mathrm{e}^{\mathrm{i}\epsilon t/\hbar}
G_\mathrm{d}^<(t, X)\nonumber\\
&=\mathrm{i}\hbar^2\frac{\sum_a\Gamma_a
f_a(\epsilon)}{[\epsilon-\mathcal{E}_\mathrm{d}(X)]^2+(\hbar\Gamma/2)^2},\\
G_\mathrm{d}^>(\epsilon, X)&=
\int\mathrm{d}t\,\mathrm{e}^{\mathrm{i}\epsilon t/\hbar}
G_\mathrm{d}^>(t, X)\nonumber\\
&=-\mathrm{i}\hbar^2\frac{\sum_a\Gamma_a\left[1-
f_a(\epsilon)\right]}{[\epsilon-\mathcal{E}_\mathrm{d}(X)]^2+(\hbar\Gamma/2)^2}.
\end{align}
\end{subequations}
In Eq.\ \eqref{eq:GF}, we defined the Fermi function in lead $a$, 
\begin{equation}
f_a(\epsilon)=\frac{1}{\exp{([\epsilon-\mu_a]/k_\mathrm{B}T)+1}}, 
\end{equation}
the effective dot energy for fixed $X$, 
\begin{equation}
\label{eq:ed}
\mathcal{E}_\mathrm{d}(X)=\epsilon_\mathrm{d}-e\bar V_\mathrm{g}+F_\mathrm{e}X, 
\end{equation}
and the hybridization $\Gamma=\sum_a\Gamma_a$ with
$\Gamma_a=2\pi|t_a|^2\nu/\hbar$. Here, $\nu$ is the density of states in the
leads, assumed to be constant.

The Fokker-Planck equation \eqref{eq:FP} can be conveniently rewritten using 
the relevant time, length and energy scales of the problem, i.e., the
vibrational frequency for vanishing compression force $\omega_0$ [see Eq.\
\eqref{eq:omega}], 
the polaronic shift $\ell=F_\mathrm{e}/m\omega_0^2$ and the corresponding energy $E_\mathrm{E}^0=F_\mathrm{e}\ell$,
respectively.
Introducing the dimensionless position, momentum and time, $x=X/\ell$,
$p=P/m\omega_0\ell$, and $\tau=\omega_0 t$, respectively, one obtains for
Eq.\ \eqref{eq:FP}
\begin{equation}
\label{eq:FP_scaled}
\partial_\tau\mathcal{P}=-p\partial_x\mathcal{P}-f_\mathrm{eff}(x)\partial_p\mathcal{P}
+\gamma(x)\partial_p(p\mathcal{P})+\frac{d(x)}{2}\partial_p^2\mathcal{P}.
\end{equation}
The
scaled effective force 
\begin{equation}
\label{eq:f_eff}
f_\mathrm{eff}(x)=\delta x-\tilde\alpha x^3-n_0(x)
\end{equation}
is expressed in terms of the reduced compression force $\delta=F/F_\mathrm{c}-1$ and the scaled
anharmonicity $\tilde\alpha=\alpha\ell^4/E_\mathrm{E}^0$.
In the low-temperature limit (or equivalently the
strong coupling limit) $k_\mathrm{B}T\ll\hbar\Gamma$, i.e., the cotunneling
regime where resonant transport is
fully coherent, the occupation of the dot for fixed deflection $x$, the (scaled) current-induced dissipation
and fluctuation read \cite{pisto08_PRB, nocer11_PRB}
\begin{align}
\label{eq:n0}
n_0(x)=&\;\frac 12+\sum_a\frac{\gamma_a}{\pi}
\arctan{\left(\frac{\tilde\mu_a+v_\mathrm{g}-x}{\tilde\Gamma/2}\right)},\\
\label{eq:gamma}
\gamma(x)=&\;\frac{\tilde\omega_0\tilde\Gamma^2}{4\pi}\sum_a
\frac{\gamma_a}{{[(\tilde\mu_a+v_\mathrm{g}-x)^2+(\tilde\Gamma/2)^2]}^2},\\
\label{eq:d}
d(x)=&\;\frac{2\gamma_\mathrm{L}\gamma_\mathrm{R}\tilde\omega_0}{\pi\tilde\Gamma}
\sum_{a\neq b}
\theta(\tilde\mu_a-\tilde\mu_b)
\nonumber\\
&\times
\left[\arctan{z}+\frac{z}{1+z^2}
\right]^{\frac{\tilde\mu_a+v_\mathrm{g}-x}{\tilde\Gamma/2}}_{\frac{\tilde\mu_b+v_\mathrm{g}-x}{\tilde\Gamma/2}},
\end{align}
respectively. We introduced the notation $\gamma_a=\Gamma_a/\Gamma$,
$v_\mathrm{g}=(e\bar V_\mathrm{g}-\epsilon_\mathrm{d})/E_\mathrm{E}^0$, 
$\tilde \mu_\mathrm{L}=-\tilde\mu_\mathrm{R}=v/2$ with $v=eV/E_\mathrm{E}^0$, 
$\tilde\omega_0=\hbar\omega_0/E_\mathrm{E}^0$, and
$\tilde\Gamma=\hbar\Gamma/E_\mathrm{E}^0$.
Solving for the stationary solution
of the Fokker-Planck equation
\eqref{eq:FP_scaled}, $\partial_\tau\mathcal{P}_\mathrm{st}=0$, one can calculate the average current through the device
according to 
\begin{equation}
\label{eq:I}
I=\int\mathrm{d}x\mathrm{d}p\; \mathcal{I}(x)\mathcal{P}_\mathrm{st}(x,p), 
\end{equation}
with 
\begin{align}
\label{eq:I(x)}
\mathcal{I}(x)=&\;\frac{e\Gamma\gamma_\mathrm{L}\gamma_\mathrm{R}}{\pi}
\left[
\arctan{\left(\frac{\tilde\mu_\mathrm{L}+v_\mathrm{g}-x}{\tilde\Gamma/2}\right)}
\right.
\nonumber\\
&\left.-\arctan{\left(\frac{\tilde\mu_\mathrm{R}+v_\mathrm{g}-x}{\tilde\Gamma/2}\right)}
\right]
\end{align}
the quasi-stationary current for fixed $x$.

Before we proceed, we notice that 
for the experiment of Steele \textit{et al.} \cite{steel09_Science} on suspended
carbon nanotubes, it was estimated in
Ref.\ \onlinecite{weick11_PRB} that $E_\mathrm{E}^0\simeq\unit[5]{\mu eV}$ and
$\tilde\alpha\simeq 10^{-10}$, yielding $\tilde\Gamma\simeq10$,
$\tilde\omega_0\simeq0.1$, and a reduced temperature $\tilde
T=k_\mathrm{B}T/E_\mathrm{E}^0\simeq1$. This experiment is thus in the
low-temperature adiabatic regime which is investigated here where 
$\tilde\omega_0\ll\tilde T\ll\tilde\Gamma$, and where cotunneling might play a
role on the transport characteristics. As for the similar experiments by
Lassagne \textit{et al.} \cite{lassa09_Science},
$E_\mathrm{E}^0\simeq\unit[5]{\mu eV}$ and
$\tilde\alpha\simeq10^{-8}$, \cite{weick11_PRB} such that we have $\tilde\Gamma\simeq 0.5$,
$\tilde\omega_0\simeq10^{-2}$ and $\tilde T\simeq10^3$.
The experiment of Ref.\ \onlinecite{lassa09_Science} is thus clearly in the high-temperature
regime $\tilde\omega_0\ll\tilde\Gamma\ll\tilde T$ where sequential transport
dominates over cotunneling and which has been investigated in detail in Ref.\
\onlinecite{weick11_PRB}.

\section{Transport characteristics of the device}
\label{sec:IV}
We now turn to the study of the transport characteristics of the device of Fig.\
\ref{fig:setup} with the approach detailed in Sec.\ \ref{sec:FP}.
We start in Sec.\ \ref{sec:mean-field} with a mean-field approximation that
neglects the current-induced dissipation \eqref{eq:gamma} and fluctuation
\eqref{eq:d}. We then consider the full non-equilibrium dynamics of the device in
Sec.\ \ref{sec:noneq} before we investigate the effects of a finite quality factor of
the nanobeam in Sec.\ \ref{sec:Q}.

\subsection{Mean-field approximation}
\label{sec:mean-field}
It is first instructive to consider the limit $\tilde\omega_0\rightarrow0$ to
understand the transport characteristics of the system.
Within this limit, the current-induced dissipation \eqref{eq:gamma} and fluctuation 
\eqref{eq:d} can be neglected as compared to the effective force \eqref{eq:f_eff}
entering the Fokker-Planck equation \eqref{eq:FP_scaled}. In that case, any
infinitesimally small extrinsic damping mechanism will, at zero temperature, localize the system at a
deflection $x$ corresponding to the global minimum $x_\mathrm{m}$ of the effective potential 
\begin{align}
\label{eq:v_eff}
v_\mathrm{eff}(x)=&-\frac{\delta x^2}{2}+\frac{\tilde\alpha x^4}{4}+\frac{x}{2}
\nonumber\\
&-\sum_a\frac{\gamma_a}{\pi}
\Bigg[
\left(\tilde\mu_a+v_\mathrm{g}-x\right)
\arctan{\left(\frac{\tilde\mu_a+v_\mathrm{g}-x}{\tilde\Gamma/2}\right)}
\nonumber\\
&-\frac{\tilde\Gamma}{4}\ln{\left((\tilde\mu_a+v_\mathrm{g}-x)^2+(\tilde\Gamma/2)^2\right)}
\Bigg]
\end{align}
associated to the effective force \eqref{eq:f_eff}. 
Within this mean-field
approximation, the stationary
solution of the Fokker-Planck equation \eqref{eq:FP_scaled} is given 
by $\mathcal{P}_\mathrm{st}(x,p)=\delta(x-x_\mathrm{m})\delta(p)$ such that the
current \eqref{eq:I} reduces to $I=\mathcal{I}(x_\mathrm{m})$, with
$\mathcal{I}(x)$ the quasi-stationary current of Eq.\ \eqref{eq:I(x)}.

The mechanism of the classical current blockade is encapsulated in the
effective potential \eqref{eq:v_eff} and in the quasi-stationary current
\eqref{eq:I(x)}. Indeed, 
the effective potential \eqref{eq:v_eff} can, depending on the parameters, show up
to five metastable minima. These positions $x$ can 
either correspond to an effective dot energy level \eqref{eq:ed} located within
the bias window, $\mathcal{E}_\mathrm{d}(X)\leqslant e|V|/2$, where sequential
tunneling dominates and where the current is maximal (``conducting" minima), either to positions such
that $\mathcal{E}_\mathrm{d}(X)> e|V|/2$ where cotunneling dominates and the
current is algebraically suppressed on a scale given by the hybridization $\Gamma$
(``blocked" minima). For a given gate voltage, the system passes
for increasing bias voltage from a stable position $x_\mathrm{m}$ corresponding
to the cotunneling (blocked) region to a stable position where sequential
tunneling dominates (conducting region), thus defining a bias voltage below which
current is suppressed.

\begin{widetext}
\subsubsection{Limit of vanishing hybridization to the leads.}
We start by considering the limit
$\tilde \Gamma\rightarrow0$ such that the dot occupation \eqref{eq:n0} simplifies
to 
\begin{equation}
\label{eq:n_0_Gamma_0}
n_0(x)=
\begin{cases}
1, &
\displaystyle
x<v_\mathrm{g}-\frac{|v|}{2},\\
\gamma_\mathrm{L}\Theta(v)+\gamma_\mathrm{R}\Theta(-v), &
\displaystyle
v_\mathrm{g}-\frac{|v|}{2}\leqslant x\leqslant v_\mathrm{g}+\frac{|v|}{2},\\
0, & 
\displaystyle
x>v_\mathrm{g}+\frac{|v|}{2}, 
\end{cases}
\end{equation}
where $\Theta(z)$ is the Heaviside step function.
In that case, the dynamical equilibrium condition 
$f_\mathrm{eff}(x)=0$ and $\mathrm{d}f_\mathrm{eff}/\mathrm{d}x<0$ with $f_\mathrm{eff}(x)$
the effective force \eqref{eq:f_eff} determining the metastable positions of the
system can be solved analytically. We find that the
regions where current can flow are
delineated in the $v$-$v_\mathrm{g}$ plane by straight lines $v\sim\pm
2v_\mathrm{g}$. For $v>0$, and to first order in the small parameter
$\tilde\alpha^{1/3}/|\delta|$ ($|\delta|/\tilde\alpha^{1/3}$) far (close) to the
buckling instability, 
the apex of the resulting conducting region defines a gap
$\Delta_v$ below which transport is blocked, 
\begin{equation}
\label{eq:gap}
\Delta_v=
\begin{cases}
\displaystyle
\vspace{.2truecm}
-\frac{1}{2\delta}, & -\delta\gg{\tilde\alpha}^{1/3},\\
\displaystyle
\vspace{.2truecm}
\frac{1}{4\delta}, & \delta\gg{\tilde\alpha}^{1/3},\\
\displaystyle
\vspace{.2truecm}
\frac{1}{2{\tilde\alpha}^{1/3}\gamma_\mathrm{R}}
\left[
\frac 32 \left(1-\gamma_\mathrm{L}^{1/3}\right)
+\frac{\delta}{\tilde\alpha^{1/3}}\left(1-\gamma_\mathrm{L}^{-1/3}\right)
\right],
& |\delta|\ll{\tilde\alpha}^{1/3}. 
\end{cases}
\end{equation}
The gate voltage at which one obtains the minimal threshold \eqref{eq:gap} reads
\begin{equation}
\label{eq:vg}
v_\mathrm{g}^\mathrm{min}=
\begin{cases}
\displaystyle
\vspace{.2truecm}
\frac{\gamma_\mathrm{L}+1/2}{2\delta}, & -\delta\gg{\tilde\alpha}^{1/3},\\
\displaystyle
\vspace{.2truecm}
-\frac{\gamma_\mathrm{L}+1/2}{4\delta}-\sqrt{\frac{\delta}{\tilde\alpha}}, &
\delta\gg{\tilde\alpha}^{1/3},\\
\displaystyle
\vspace{.2truecm}
-\frac{1}{8{\tilde\alpha}^{1/3}\gamma_\mathrm{R}}
\Bigg\{
3 \left[1+\gamma_\mathrm{L}^{1/3}(\gamma_\mathrm{R}-\gamma_\mathrm{L})\right]
+\frac{2\delta}{\tilde\alpha^{1/3}}
\left[1+\gamma_\mathrm{L}^{-1/3}(\gamma_\mathrm{R}-\gamma_\mathrm{L})\right]
\Bigg\},
& |\delta|\ll{\tilde\alpha}^{1/3}. 
\end{cases}
\end{equation}
\end{widetext}
The case $v<0$ is easily obtained by swapping L and R in Eqs.\ \eqref{eq:gap} and
\eqref{eq:vg}.

\begin{figure*}[tb]
\includegraphics[width=1.7\columnwidth]{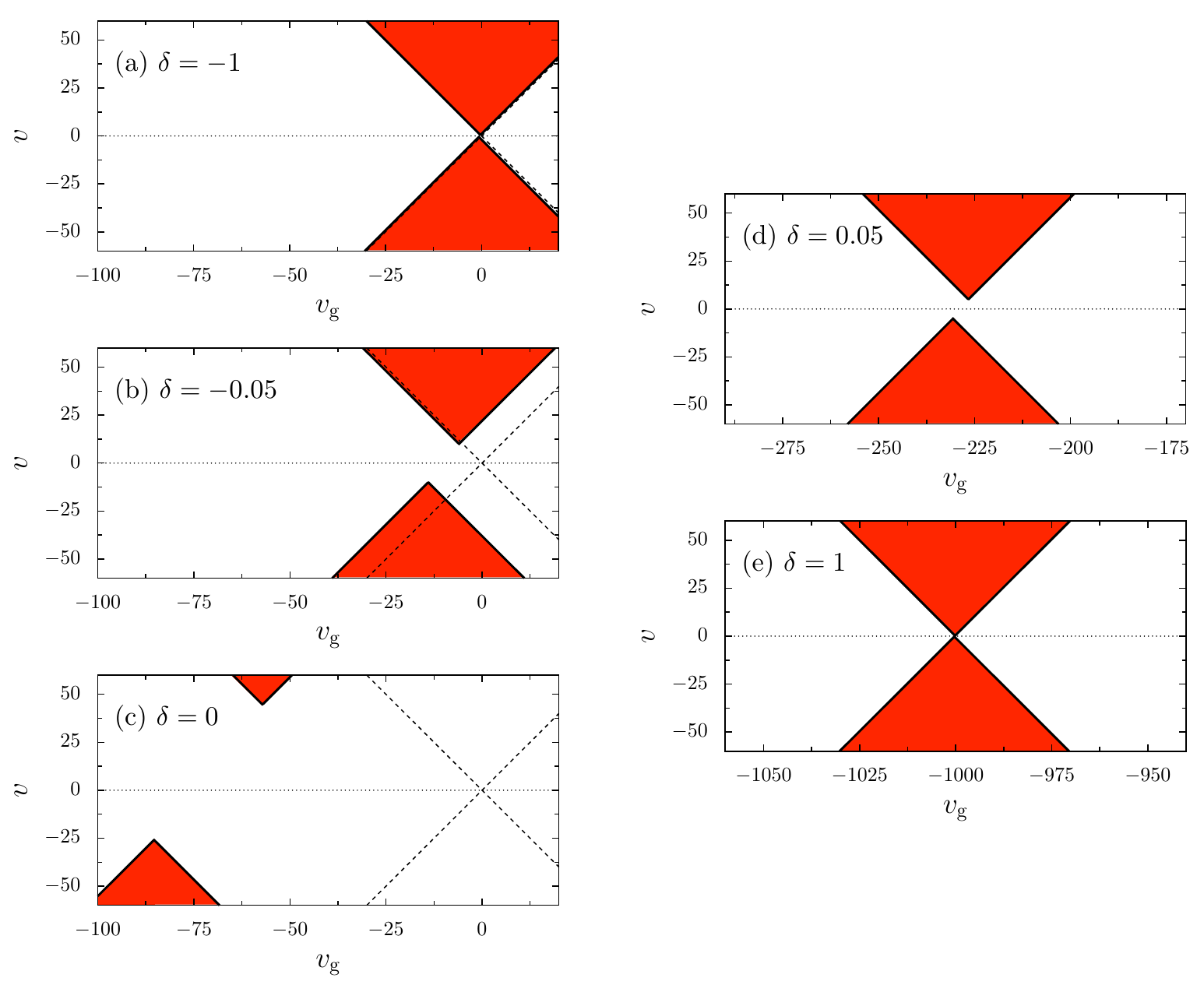}
\caption{
\label{fig:diamonds}%
(Color online)
Conducting regions at mean-field level (in red/dark gray) in the $v$-$v_\mathrm{g}$ plane for
$\tilde\Gamma\rightarrow0$. The red/dark gray and white areas correspond to a current
$|I|=e\Gamma\gamma_\mathrm{L}\gamma_\mathrm{R}$ and $I=0$, respectively. 
The scaled compression force $\delta$ increases from (a) to
(e). 
In the figure, $\gamma_\mathrm{L}=0.1$, $\gamma_\mathrm{R}=1-\gamma_\mathrm{L}=0.9$ and 
$\tilde\alpha=10^{-6}$. The dashed lines indicate the
delimitation of the conducting regions for vanishing electromechanical coupling
($F_\mathrm{e}=0$).}
\end{figure*}

The regions in the $v$-$v_\mathrm{g}$ plane where current can flow through the
nanobeam are shown in Fig.\ \ref{fig:diamonds} in red/dark gray.
As it is the case in the sequential-tunneling regime, \cite{weick11_PRB}
the current blockade
characterized by the gap \eqref{eq:gap} increases for increasing compression
force $\delta<0$ below the mechanical instability [see Figs.\ \ref{fig:diamonds}(a)
and \ref{fig:diamonds}(b)]. The gap is maximal at the Euler buckling
instability where $\delta=0$ [see Fig.\ \ref{fig:diamonds}(c)] before it
decreases above the
instability [$\delta>0$, see Figs.\ \ref{fig:diamonds}(d) and
\ref{fig:diamonds}(e)].
Since $\tilde\alpha\ll1$, \cite{weick11_PRB} $\Delta_v$ is orders of magnitude larger than away from the
mechanical instability. 
Moreover, one observes in Fig.\ \ref{fig:diamonds} a shift of the tip of the conducting regions 
for positive and negative bias voltages which is due to the asymmetry of the coupling to the
left and right leads ($\gamma_\mathrm{L}\neq\gamma_\mathrm{R}$).
Indeed, this shift which is maximal at the
buckling instability [Fig.\ \ref{fig:diamonds}(c)] 
can be traced back to the different occupations of the dot for fixed $x$
[Eq.\ \eqref{eq:n_0_Gamma_0}] for positive and negative biases when $x$ is such that the
effective energy level of the dot \eqref{eq:ed} belongs to the
sequential-tunneling region. Such an asymmetry of the conducting regions for
positive and negative bias voltages has also been observed experimentally
in suspended carbon nanotubes for quantized vibrational modes 
\cite{letur09_NaturePhysics} and is thus not specific to the classical regime
considered here. 

We conclude this section by noticing that the limit $\tilde\Gamma\rightarrow0$
presented here is quantitatively similar to the
zero-temperature mean-field limit considered in Ref.\ \onlinecite{weick11_PRB} where
sequential transport and symmetric coupling to the leads were assumed,
as the effective potential $v_\mathrm{eff}(x)$ has the same
functional form in both cases
[compare Eqs.\ \eqref{eq:gap} and \eqref{eq:vg} above for
$\gamma_\mathrm{L}=\gamma_\mathrm{R}=1/2$
with Eqs.\ (23) and (24) in Ref.\ \onlinecite{weick11_PRB}]. 
Hence, the enhancement of the current blockade
is, at mean-field level, not specific to the transport regime considered.

\subsubsection{Finite hybridization to the leads.}
\begin{figure}[tb]
\includegraphics[width=\columnwidth]{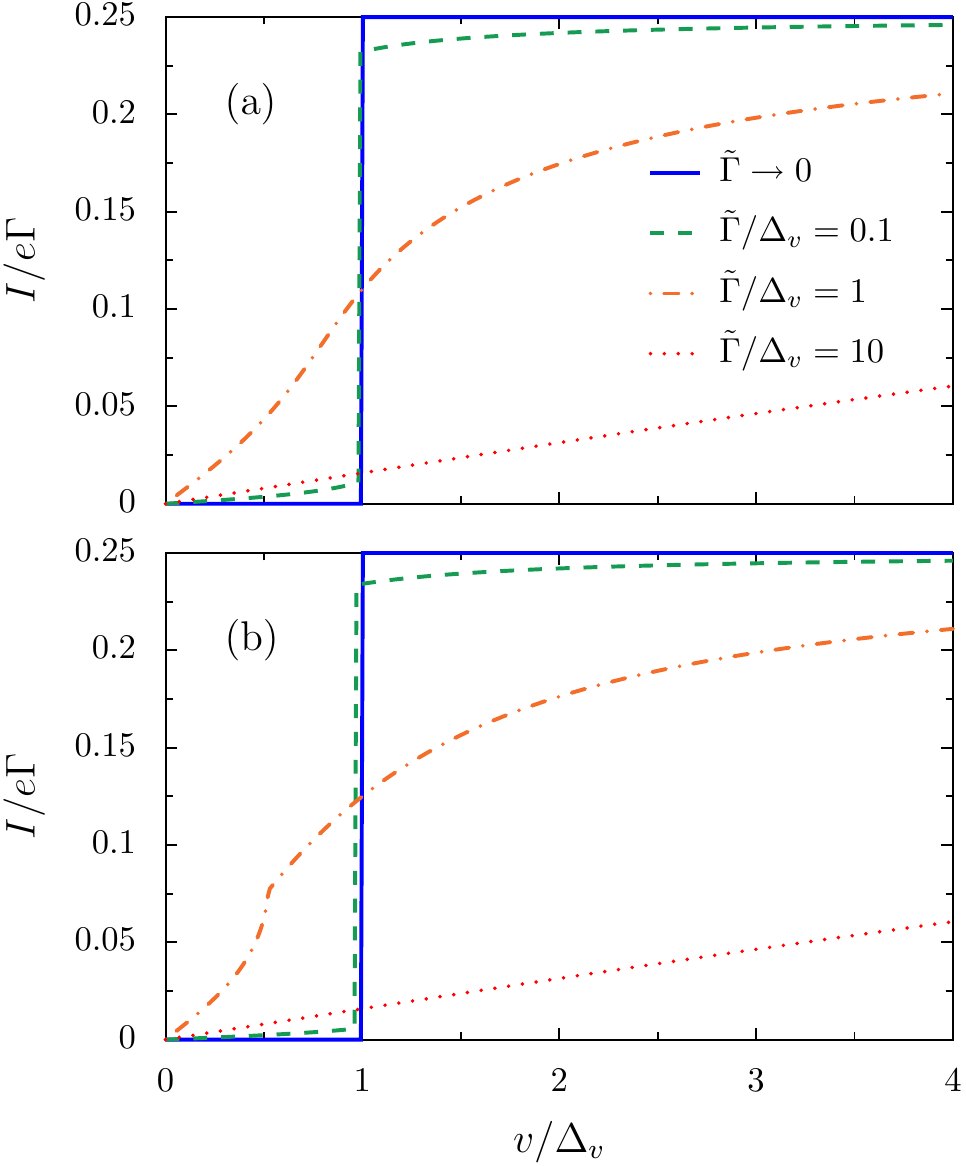}
\caption{
\label{fig:finite_gamma}%
(Color online)
Mean-field current $I$ as a function of bias $v$ scaled by the gap \eqref{eq:gap} 
for a gate voltage $v_\mathrm{g}=v_\mathrm{g}^\mathrm{min}$ corresponding to the apex of the conducting region, see
Eq.\ \eqref{eq:vg}. (a) Compression force $\delta=0$ corresponding to the
Euler instability. (b) Compression forces $|\delta|\gg\tilde\alpha^{1/3}$ far
from the buckling instability.
In the figure, 
symmetric coupling to the leads is considered
($\gamma_\mathrm{L}=\gamma_\mathrm{R}=1/2$).}
\end{figure}

At finite $\tilde\Gamma$, we search for the global minimum of the effective
potential \eqref{eq:v_eff}
numerically. The resulting average current is shown
in Fig.\ \ref{fig:finite_gamma} as a function of the bias voltage for gate
voltages corresponding to the apex of the conducting region,
$v_\mathrm{g}=v_\mathrm{g}^\mathrm{min}$ [cf.\ Eq.\ 
\eqref{eq:vg}]. 
One finds that the current behavior, plotted as a function
of $v/\Delta_v$, is similar for a given $\tilde\Gamma/\Delta_v$ at the
instability [Fig.\ \ref{fig:finite_gamma}(a)]
and away from the instability [Fig.\ \ref{fig:finite_gamma}(b)]. As the
hybridization $\tilde\Gamma$ is increased, the low-bias current blockade becomes
less pronounced since 
cotunneling becomes dominant in the blocked region ($v<\Delta_v$). For 
$\tilde\Gamma\gg\Delta_v$ (red dotted line), the blockade completely
disappears. In that limit, one recovers the usual linear-response current 
in absence of electromechanical coupling
($F_\mathrm{e}=0$), $I=4\gamma_\mathrm{L}\gamma_\mathrm{R}G_0 V$, with
$G_0=e^2/2\pi\hbar$ the conductance quantum. The results
of Fig.\ \ref{fig:finite_gamma} show (at mean-field level) that it is
advantageous to tune the system at the Euler instability, where the gap
$\Delta_v$ dramatically increases [see Eq.\ \eqref{eq:gap}] such that the
blockade is much more pronounced for a given hybridization $\tilde\Gamma$ to
the leads. 

\subsection{Non-equilibrium dynamics}
\label{sec:noneq}
We now turn to the role played by the current-induced dissipation \eqref{eq:gamma} and
fluctuations \eqref{eq:d} on the current blockade
by considering the case $\tilde\omega_0\neq0$.
To this end, we solve for the stationary solution of the Fokker-Planck equation
\eqref{eq:FP_scaled} numerically as described in Ref.\ \onlinecite{pisto08_PRB}.
In what follows, we focus on the system at the Euler instability ($\delta=0$)
where the current blockade is, at mean-field level, maximal [cf.\ Eq.\ 
\eqref{eq:gap}].
Since the stationary solution of the Fokker-Planck equation \eqref{eq:FP_scaled}
is, for $\tilde\omega_0\ll1$, independent of the actual value of
$\tilde\omega_0$ (see appendix E in Ref.\ \onlinecite{weick11_PRB} for details), we
set $\tilde\omega_0=10^{-3}$ in what follows.

\begin{figure}[tb]
\includegraphics[width=\columnwidth]{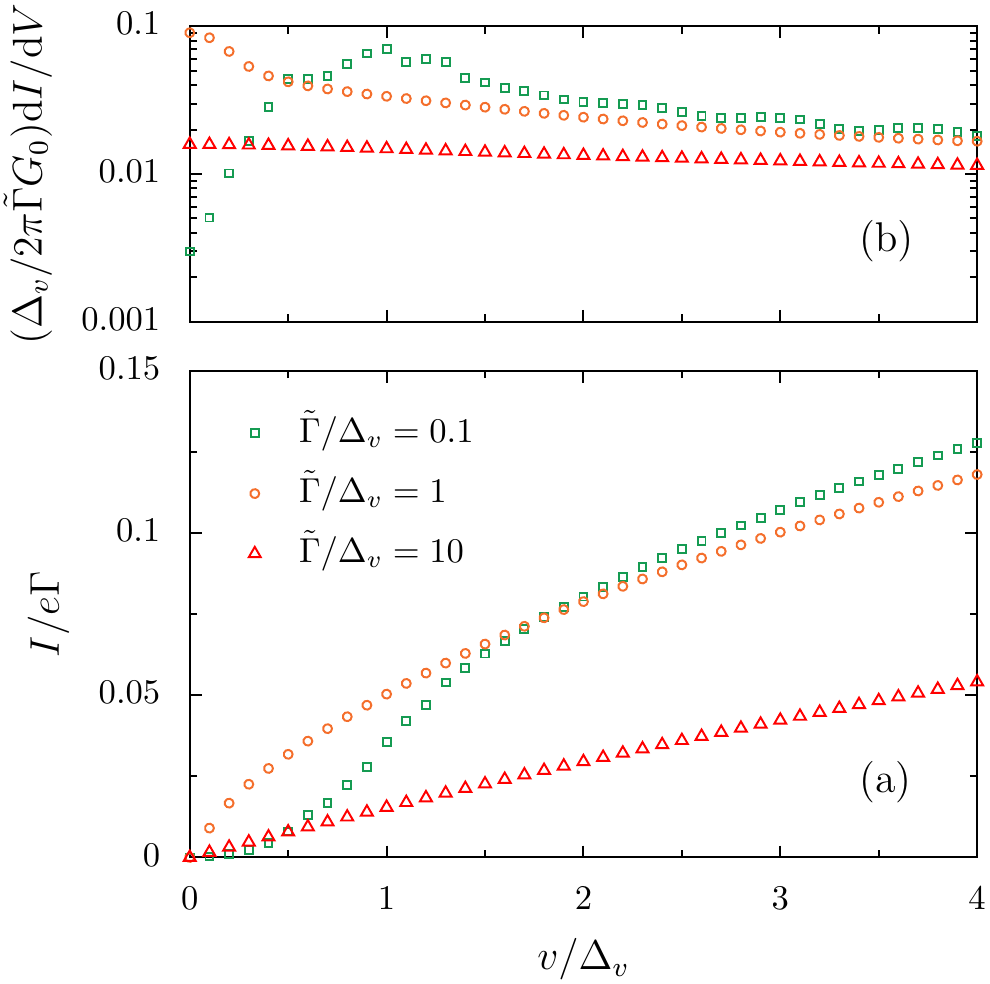}
\caption{
\label{fig:fluctuations}%
(Color online)
(a) Current $I$ including the nonequilibrium Langevin dynamics as a function of bias $v$ at the Euler buckling instability
($\delta=0$) for a gate voltage $v_\mathrm{g}=v_\mathrm{g}^\mathrm{min}$ corresponding to the apex of the
conducting region, see Eq.\ \eqref{eq:vg}.
(b) Corresponding differential conductance $\mathrm{d}I/\mathrm{d}V$, where 
$G_0=e^2/2\pi\hbar$ denotes the conductance quantum.
In the figure, 
symmetric coupling to the leads is considered
($\gamma_\mathrm{L}=\gamma_\mathrm{R}=1/2$), $\tilde\alpha=10^{-6}$ and
$\tilde\omega_0=10^{-3}$.}
\end{figure}

Our results for the average current are shown in Fig.\ \ref{fig:fluctuations}(a) for
increasing hybridization to the leads $\tilde\Gamma$ which controls the range of the
current-induced dissipation \eqref{eq:gamma} and fluctuations 
\eqref{eq:d}.
For $\tilde\Gamma$ smaller than the energy gap $\Delta_v$ (see green 
squares and orange circles in the figure), the current-induced fluctuations have a
dramatic effect on the $I$-$V$ characteristics for bias voltages larger than
$\Delta_v$ [compare with the mean-field
results of Fig.\ \ref{fig:finite_gamma}(a), green dashed and orange
dashed-dotted lines]. However, 
for $\tilde\Gamma\leqslant0.1\Delta_v$ and $v<\Delta_v$, cotunneling does not
suppress the current blockade at low bias. We expect that for smaller
$\tilde\Gamma$ (which is numerically very difficult to tackle), the current
blockade should be even more pronounced.
As close to the mechanical instability, the gap \eqref{eq:gap} is very
large, we foresee that the current blockade should be clearly visible in an
experiment even in the regime of strong coupling to the electron reservoirs
where cotunneling can be
significant.

The above results are confirmed by the behavior of the differential conductance
$\mathrm{d}I/\mathrm{d}V$ shown in Fig.\ \ref{fig:fluctuations}(b): For a
hybridization to the leads small as compared to the energy gap (green squares in
the figure), the differential conductance is clearly suppressed at low bias
voltage, while for larger $\tilde\Gamma$, the conductance acquires a finite
value which is almost constant as a function of the bias voltage [orange circles
and red triangles in Fig.\ \ref{fig:fluctuations}(b)].

\subsection{Role of extrinsic dissipation}
\label{sec:Q}
Nanoelectromechanical systems are subject to various sources of extrinsic dissipation that leads to a
finite quality factor $Q$ of the nanobeam 
that we have ignored so far. These mechanisms of
extrinsic dissipation come from localized defects at the surface of the sample,
\cite{mohan02_PRB, seoan07_EPL, seoan08_PRB}
clamping and thermoelastic losses, \cite{cleland} Ohmic losses due to the gate electrode,
\cite{seoan07_PRB} etc. Within the Caldeira-Leggett model and assuming Ohmic
(memory-free) dissipation, \cite{weiss} one can easily
incorporate these mechanisms in the Fokker-Planck equation \eqref{eq:FP_scaled}
by replacing $\gamma(x)\rightarrow\gamma(x)+\gamma_\mathrm{e}$ and
$d(x)\rightarrow d(x)+2\gamma_\mathrm{e}\tilde T$, with
$\gamma_\mathrm{e}=Q^{-1}$ the extrinsic damping constant (or inverse quality
factor) and $\tilde T=k_\mathrm{B}T/E_\mathrm{E}^0$ the (reduced) temperature.

\begin{figure}[tb]
\includegraphics[width=\columnwidth]{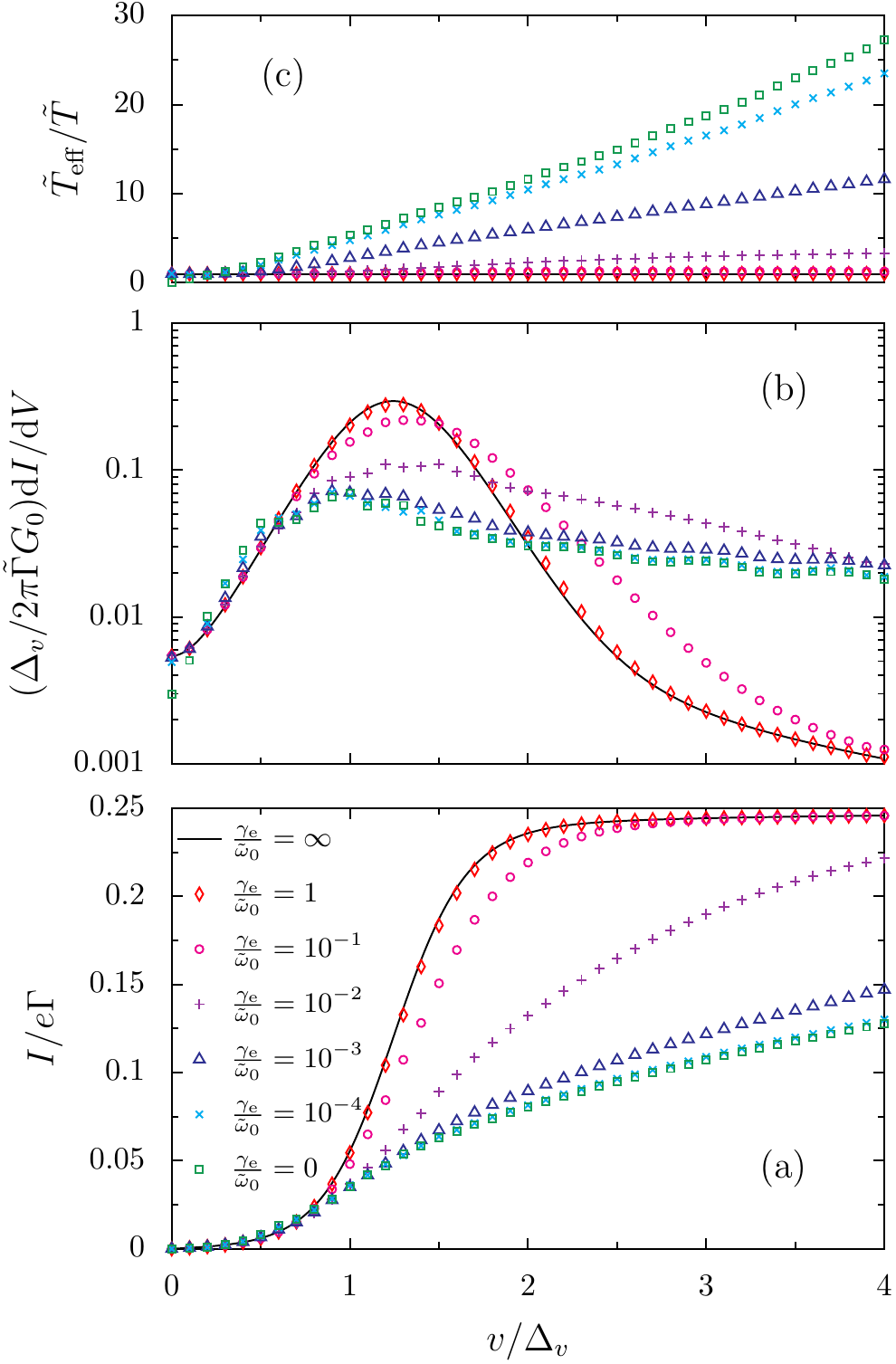}
\caption{
\label{fig:quality_factor}%
(Color online)
(a) Average current and (b) differential conductance as a function of the bias voltage at the Euler buckling
instability ($\delta=0$) for various values of the extrinsic damping constant
$\gamma_\mathrm{e}$. 
(c) Corresponding effective temperature $\tilde T_\mathrm{eff}$ defined in
Eq.\ \eqref{eq:T_eff}.
In the figure, $\tilde\alpha=10^{-6}$,
$\tilde\omega_0=10^{-3}$, $\tilde\Gamma/\Delta_v=0.1$, $\tilde
T/\Delta_v=0.05$, $\gamma_\mathrm{L}=\gamma_\mathrm{R}=1/2$, and the gate
voltage corresponds to the apex of the conducting region,  
see Eq.\ \eqref{eq:vg}.}
\end{figure}

Numerical results for various values of the
ratio $\gamma_\mathrm{e}/\tilde\omega_0$ 
are shown for the current in Fig.\ \ref{fig:quality_factor}(a) and for the
corresponding differential conductance in Fig.\ \ref{fig:quality_factor}(b). 
The parameter $\gamma_\mathrm{e}/\tilde\omega_0$ 
measures the relative importance of the two sources of fluctuations, i.e.,
thermal vs.\ current-induced [cf.\ Eqs.\ \eqref{eq:gamma} and \eqref{eq:d}], from
vanishing extrinsic dissipation [green squares in Figs.\ \ref{fig:quality_factor}(a)
and \ref{fig:quality_factor}(b)] to
purely thermal fluctuations (solid line in the figure), where the probability
$\mathcal{P}_\mathrm{st}(x,p)$ entering Eq.\ \eqref{eq:I}
obeys a Boltzmann distribution at temperature $\tilde T$.
We can conclude from the results of Figs.\ \ref{fig:quality_factor}(a) and 
\ref{fig:quality_factor}(b) that the larger
the extrinsic dissipation (or the smaller the quality factor), the more pronounced
is the current blockade. As it is the case in the 
sequential-tunneling regime, \cite{weick11_PRB} the extrinsic dissipation helps
localizing the system at phase-space points corresponding to the metastable minima of the
effective potential \eqref{eq:v_eff}, rendering the current blockade sharper.
The same conclusion holds for quantized vibrations, where the Franck-Condon
blockade is more pronounced for fast equilibration of the vibronic mode.
\cite{koch05_PRL, koch06_PRB}

The results for the average current in Fig.\ \ref{fig:quality_factor}(a) and
for the differential conductance in Fig.\ \ref{fig:quality_factor}(b) can be 
interpreted in terms of an effective temperature 
\begin{equation}
\label{eq:T_eff}
\tilde T_\mathrm{eff}=\frac{\langle d\rangle/2+\gamma_\mathrm{e}\tilde
T}{\langle\gamma\rangle+\gamma_\mathrm{e}}
\end{equation}
defined in analogy with the
fluctuation-dissipation theorem, \cite{zwanzig} where
\begin{equation}
\langle\gamma\rangle=\int\mathrm{d}x\mathrm{d}p\;\gamma(x)\mathcal{P}_\mathrm{st}(x, p)
\end{equation}
and 
\begin{equation}
\langle d\rangle=\int\mathrm{d}x\mathrm{d}p\;d(x)\mathcal{P}_\mathrm{st}(x, p)
\end{equation}
are the averages over phase-space of the current-induced dissipation
\eqref{eq:gamma}
and fluctuation \eqref{eq:d}, respectively. As one can see in 
Fig.\ \ref{fig:quality_factor}(c), $\tilde T_\mathrm{eff}$ almost equals the
electronic temperature $\tilde T$ for inverse quality factors $\gamma_\mathrm{e}$
of the order of $\tilde\omega_0$, thus explaining why the current and
conductance behavior are
similar to the case where only thermal fluctuations are present [compare
red diamonds and magenta circles with the solid line in Figs.\ \ref{fig:quality_factor}(a)
and \ref{fig:quality_factor}(b)]. For
smaller $\gamma_\mathrm{e}$ (larger quality factors $Q$), the effective
temperature becomes significantly larger than $\tilde T$ and increases with
increasing bias voltage. Hence, the mechanical system fluctuates more in phase-space and
switches back and forth between minima of the effective potential
\eqref{eq:v_eff} corresponding to regions where the current at fixed $x$, Eq.\ 
\eqref{eq:I(x)},
is suppressed (cotunneling/``blocked" regions) or enhanced
(sequential tunneling/``conducting"
regions), thus reducing the average current $I$ and rendering the
current blockade less pronounced.

We conclude this section on the transport characteristics of the device by
noticing that the features of the classical current blockade in the vicinity of
a mechanical instability are qualitatively the same in the sequential-tunneling
and cotunneling transport regimes. 
In the former case, the electronic temperature defines the relevant energy scale
below which the current blockade is observable, \cite{weick11_PRB} while in the
latter case, it is the
hybridization to the electron reservoirs that plays a similar role.
The enhancement of the current blockade at the Euler instability is thus a
universal phenomenon that does not depend on the transport regime one considers.

\section{Backaction of the current flow on the nanoresonator}
\label{sec:backaction}

\begin{figure*}[tb]
\includegraphics[width=2\columnwidth]{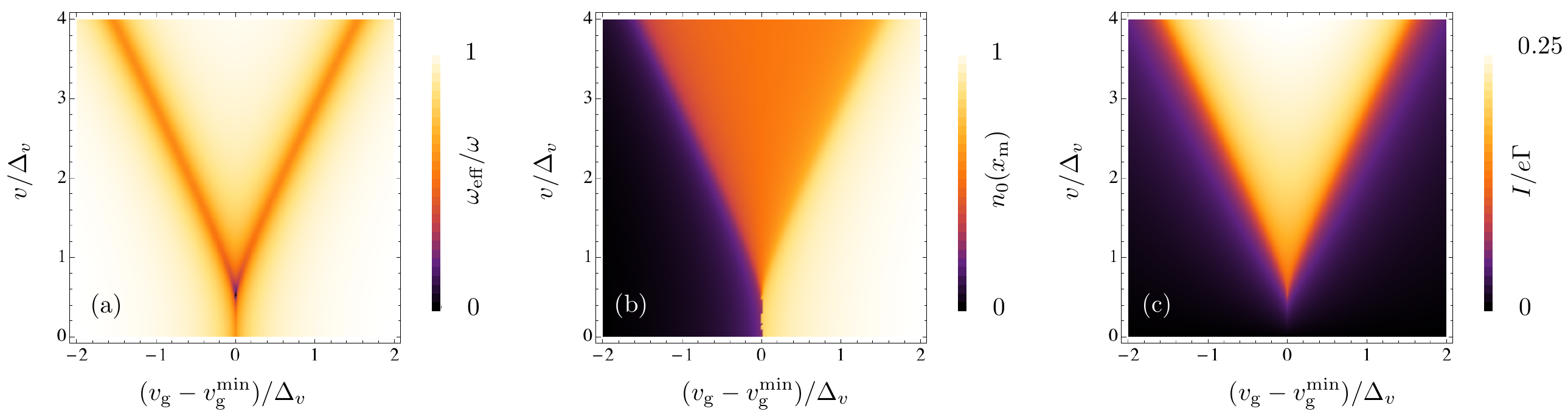}
\caption{
\label{fig:frequency_diamond}%
(Color online)
(a) Effective frequency $\omega_\mathrm{eff}$, (b) average occupation of the dot and (c) average current $I$
as a function of bias voltage $v$ and gate voltage $v_\mathrm{g}$ 
for compression forces far below the
Euler instability ($-\delta\gg\tilde\alpha^{1/3}$).
In the figure, $\tilde\Gamma=\Delta_v$ and
$\gamma_\mathrm{L}=\gamma_\mathrm{R}=1/2$.}
\end{figure*}

A characteristics of nanoelectromechanical systems is the backaction that the current flow exerts on the
mechanical part of the device. The current-induced force $-n_0(x)$ in the effective
force \eqref{eq:f_eff} can indeed renormalize the nanobeam frequency. Since this
current-induced force depends on the bias and gate voltages through the average occupation of the dot for
fixed position $x$, one expects a strong signature
of electronic transport on the nanobeam frequency.

In what follows, we quantify the current-induced effective frequency
$\omega_\mathrm{eff}$ of the nanobeam
by the curvature of the effective potential \eqref{eq:v_eff} at its global
minimum $x_\mathrm{m}$,
$\omega_\mathrm{eff}=\omega_0\sqrt{v_\mathrm{eff}''(x_\mathrm{m})}$. Using
Eqs.\ \eqref{eq:f_eff} and \eqref{eq:n0}, we have
\begin{equation}
\label{eq:omega_eff}
\omega_\mathrm{eff}=\omega_0\sqrt{-\delta+3\tilde\alpha
x_\mathrm{m}^2-\left.\frac{\partial n_0}{\partial
v_\mathrm{g}}\right|_{x_\mathrm{m}}}.
\end{equation}
Since $\partial n_0/\partial v_\mathrm{g}>0$, this expression shows that the
nanobeam frequency will be significantly lowered when the average number of
electrons on the dot varies as a function of gate voltage. 
This is illustrated in Figs.\ \ref{fig:frequency_diamond}(a) and
\ref{fig:frequency_diamond}(b) where we show
$\omega_\mathrm{eff}$ and the average number of electrons on the dot $n_0$ 
as a function of bias $v$ and gate voltage $v_\mathrm{g}$, respectively, for
intermediate hybridization to the leads $\tilde\Gamma=\Delta_v$.
For bias voltages $v\lesssim\Delta_v$, the average occupation of the dot in
Fig.\ \ref{fig:frequency_diamond}(b) jumps from 0 to 1 around a gate voltage
corresponding to Eq.\ \eqref{eq:vg} and on a scale given by the coupling to the leads
$\tilde\Gamma$, resulting in a renormalization of the nanobeam frequency along
the line $v_\mathrm{g}=v_\mathrm{g}^\mathrm{min}$, see Fig.\ 
\ref{fig:frequency_diamond}(a). For $v\gtrsim\Delta_v$, the frequency is mostly
renormalized for gate voltages
corresponding to the borders between the sequential tunneling region (where
$n_0\simeq 1/2$) and the cotunneling regions (where $n_0\simeq0$ or 1). This is
further illustrated in Fig.\ 
\ref{fig:frequency_diamond}(c) which shows the corresponding average
current flowing through the nanobeam. 
The results of Fig.\ \ref{fig:frequency_diamond} are obtained for compression
forces $-\delta\gg\tilde\alpha^{1/3}$, i.e., far below the Euler instability.
Far above the instability ($\delta\gg\tilde\alpha^{1/3}$), the results of Fig.\
\ref{fig:frequency_diamond} remain unchanged, except that one has to replace
$\omega_\mathrm{eff}/\omega$ by $\omega_\mathrm{eff}/\sqrt{2}|\omega|$ in Fig.\ 
\ref{fig:frequency_diamond}(a) since 
in the buckled state, the vibrational
frequency squared in absence of a current flow is twice as much as in the flat
state, see Eq.\ \eqref{eq:omega_eff}.

\begin{figure}[b]
\includegraphics[width=\columnwidth]{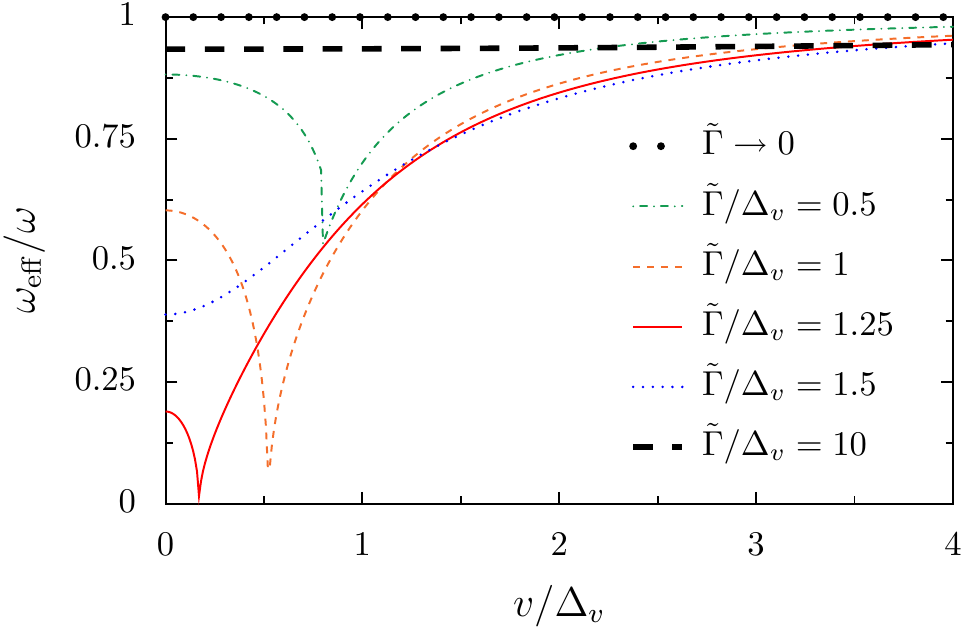}
\caption{
\label{fig:frequency_shift}%
(Color online)
Effective frequency $\omega_\mathrm{eff}$ of the nanobeam far below the
Euler instability ($-\delta\gg\tilde\alpha^{1/3}$)
as a function of bias voltage at 
the apex of the conducting region, $v_\mathrm{g}=v_\mathrm{g}^\mathrm{min}$.
In the figure, $\gamma_\mathrm{L}=\gamma_\mathrm{R}=1/2$.}
\end{figure}

In Fig.\ \ref{fig:frequency_shift}, we study the influence of the hybridization
$\tilde\Gamma$ on the current-induced frequency shift.
We show in the figure the effective frequency
$\omega_\mathrm{eff}$ as a
function of the bias voltage for gate voltages corresponding to the apex of the
conducting region [cf.\ Eq.\ \eqref{eq:vg}], for compression forces far below the
Euler instability ($-\delta\gg\tilde\alpha^{1/3}$), and for various values of
the hybridization $\tilde\Gamma$ to the leads. 
It is remarkable that for
$\tilde\Gamma\rightarrow0$ (thick black dotted line in Fig.\ \ref{fig:frequency_shift}), 
$\omega_\mathrm{eff}=\omega$, i.e., there is no effect of the
current flow on the vibrational frequency of the beam [since in that limit,
the curvatures of the effective potential \eqref{eq:v_eff} is the same in all
three metastable minima], while the current blockade is the most pronounced [see
blue solid line in Fig.\ \ref{fig:finite_gamma}(b)]. 
For $\tilde\Gamma\gg\Delta_v$, the effective frequency
$\omega_\mathrm{eff}/\omega\simeq1-2\Delta_v/\pi\tilde\Gamma$ is slightly reduced
as compared to its value without a current flow and is independent of the bias
voltage (see thick dashed line in Fig.\ \ref{fig:frequency_shift}). 
Only for $\tilde
\Gamma\sim\Delta_v$ (see thin lines in Fig.\ \ref{fig:frequency_shift}) the
beam experiences large bias-dependent frequency shifts due to the current flow, that can
reach up to almost $\unit[100]{\%}$ (see red solid line in the figure). This is
in stark contrast with the behavior of the current
[Fig.\ \ref{fig:finite_gamma}(b)], where the low-bias current blockade already almost
disappears for intermediate values of the hybridization [see orange dashed-dotted
line in Fig.\ \ref{fig:finite_gamma}(b)]. 

The results above show that, although the current blockade might
not be experimentally accessible far below (or above) the Euler instability for
intermediate hybridization to the leads, the
current-induced frequency shift is considerably easier to detect. This was recently
demonstrated experimentally on suspended carbon nanotubes,
\cite{steel09_Science, lassa09_Science} where despite the fact that the current
blockade could not be detected due to the relatively weak electromechanical
coupling encountered in experiments, a clear current-induced frequency shift was measured.

\section{Conclusion}
\label{sec:ccl}
In conclusion, we have analyzed the role played by cotunneling, relevant in the
transport regime of strong hybridization to the electron reservoirs, on the
low-bias current blockade in suspended single-electron transistors. We have
shown that the dramatic enhancement at the Euler buckling instability of the bias voltage below which transport
is suppressed predicted in Ref.\ \onlinecite{weick11_PRB} is not ruled out by cotunneling currents. 
The same conclusion holds in molecular devices for the Franck-Condon blockade in the
quantum regime.
\cite{koch06_PRB}
While in the sequential-tunneling regime, the temperature sets the limit of
observability of the current blockade, it is the hybridization to the leads that
plays a role qualitatively similar to the temperature when transport is fully
coherent.
The mechanism of the classical current blockade and of its enhancement in the
vicinity of a mechanical instability is thus a universal feature of
capacitively-coupled nanoelectromechanical systems, which does not qualitatively
depend on the transport regime.

We have further studied the effect of backaction forces on
the mechanical properties of the nanobeam.
We have shown that the mechanical frequency can be strongly renormalized
as a function of the bias voltage due to the force that is exerted by the
current flow on the nanobeam.
This effect might be considerably easier to observe than the current blockade when the
system is far from the Euler buckling instability.

\begin{acknowledgments}
We acknowledge Felix von Oppen, Fabio Pistolesi
and Dietmar Weinmann for helpful comments and discussions. 
\end{acknowledgments}


\end{document}